# Contraction of Online Response to Major Events

**Michael Szell[1]*, Sébastian Grauwin[1,2,3], Carlo Ratti[1]**

1 Senseable City Laboratory, Massachusetts Institute of Technology, Cambridge, Massachusetts, United States of America, 2 Rhône Alpes Institute of Complex Systems, Université de Lyon, Lyon, France, 3 Computer science Laboratory (LIP) - UMR 5668, CNRS, INRIA, Ecole Normale Supérieure de Lyon, Lyon, France

## Abstract

Quantifying regularities in behavioral dynamics is of crucial interest for understanding collective social events such as panics or political revolutions. With the widespread use of digital communication media it has become possible to study massive data streams of user-created content in which individuals express their sentiments, often towards a specific topic. Here we investigate messages from various online media created in response to major, collectively followed events such as sport tournaments, presidential elections, or a large snow storm. We relate content length and message rate, and find a systematic correlation during events which can be described by a power law relation—the higher the excitation, the shorter the messages. We show that on the one hand this effect can be observed in the behavior of most regular users, and on the other hand is accentuated by the engagement of additional user demographics who only post during phases of high collective activity. Further, we identify the distributions of content lengths as lognormals in line with statistical linguistics, and suggest a phenomenological law for the systematic dependence of the message rate to the lognormal mean parameter. Our measurements have practical implications for the design of micro-blogging and messaging services. In the case of the existing service Twitter, we show that the imposed limit of 140 characters per message currently leads to a substantial fraction of possibly dissatisfying to compose tweets that need to be truncated by their users.

**Funding:** Sebastian Grauwin acknowledges financial support from Ericsson's "Signature of Humanity" fellowship. Further supporters of Senseable City Laboratory are: the National Science Foundation, the AT&T Foundation, the Rockefeller Foundation, the MIT SMART program, the MIT CCES program, Audi Volkswagen, BBVA, The Coca Cola Company, Ericsson, Expo 2015, Ferrovial, GE, and all the members of the MIT Senseable City Lab Consortium. The funders had no role in study design, data collection and analysis, decision to publish, or preparation of the manuscript.

**Competing Interests:** The authors received funding from commercial sources (Ericsson, Audi Volkswagen, BBVA, The Coca Cola Company, Ericsson, Expo 2015, Ferrovial, and GE). This does not alter the authors' adherence to all the PLOS ONE policies on sharing data and materials.

* E-mail: mszell@mit.edu

## Introduction

One of the great challenges in the social sciences is the measurement and quantification of the hidden, statistical rules behind our individual decisions and the link from single to collective actions. First crucial steps in this process are the collection and relation of data about stimulus, sensation, and response. Here we exploit the recent widespread use of online social media and the availability of data on collective online reactions to wide-ranging stimuli to uncover quantitative principles in large-scale behavioral response [1]. A growing body of communication and complex network research is making use of microblogging services such as *Twitter* and of comparable online media due to their often publicly available, massive number of user-generated messages. The majority of these studies focuses their analysis on either ”macroscopic” properties only, usually on the volume of the stream (number of messages per unit time) and its temporal dynamics, or on ”microscopic” processes only such as burst-like individual activities or the cascading of information through the social network via local diffusion mechanisms [2–17].

We are interested in the effects of arousal, i.e. the psychological state which implies changes in the brain increasing the ”readiness to respond to any event external or internal” [18], and how arousal is quantitatively manifested in the online response of a large number of individuals to collectively followed, important events. Physiological effects of arousal on the human brain have been studied extensively [19], but the behavioral response to emotional stimuli is a difficult to assess, largely open question. To gain statistically meaningful insights into the matter with large corpora of data, we measure the connection between length of contents and rate of messages during events of collective excitation. We find distinct regularities, using several data sets collected from the different media of 1) Twitter, 2) a popular online forum, 3) the Enron email corpus, 4) Facebook, and 5) app.net, a Twitter-like service, see Section S1 in File S1. We formulate a quantitative relation between these two features in loose analogy to the Weber-Fechner [20,21] and Stevens [22] laws of Psychophysics, suggesting a logarithmic relation $\psi \sim \log S$ or a power law relation $\psi \sim S^{\gamma}$, between subjective sensation $\psi$ and stimulus $S$. In Psychophysics, measurements are based on individual subjective reports to changes in physical stimuli, such as the perceived intensity of lifting objects with increasing weights, to quantify the relation between stimulus and sensation in individuals. However in our case, the relation applies first to emotional rather than physical stimuli and second, not explicitly to individuals but to the collective by using aggregated data, connecting the features of length of individual messages and the rate of messages – a proxy for excitation – in the system. The observation of the same relation in several different communication media hints at a universal effect and allows us to distill the essential ingredients for formulating a possibly underlying phenomenological law. We also assess the functional distribution of content lengths of messages as lognormals, connecting to statistical linguistics and the relatively understudied distribution of sentence lengths, and show that this distribution changes its shape depending on the magnitude of excitation.

## Materials and Methods

### Events and data sets

We use several large data sets of online messages collected from different media, for details on the collection process see Section S1 in File S1. Data set 1 contains around 410,000 messages from Twitter (www.twitter.com) collected during the 2012 Masters Tournament, a major championship in golf, held between April 5 and 8, 2012 at Augusta, Georgia, USA, organized by the PGA Tour (named after PGA, the Professional Golfers' Association of America). The golfer Bubba Watson won the tournament on April 8th by defeating his opponent Louis Oosthuizen in a sudden-death playoff which marks the finale of the event. Messages in Twitter are called tweets and consist of up to 140 characters. Unregistered users can read the tweets, while registered Twitter users can post tweets through the website interface, SMS, or a range of apps for mobile devices. Every registered user has a list of users she is "following", which generates a stream of tweets visible to the user. At the same time, every user has a list of followers. Each user has the possibility to create a retweet, which copies an original tweet (or retweet). In our analysis we only consider tweets but no retweets, since only the former contain original content. This data set was extracted from Twitter by querying for a number of golf-related keywords, see Section S1 in File S1.

Data set 2 includes almost 20,000 messages posted in one thread of a popular online forum, the Something Awful (SA) forums (forums.somethingawful.com), during the US presidential election night of November 6, 2012, and a smaller number of messages posted the week before the election night. The former data set, data set 2a, has a clear climax when Barack Obama is confirmed winning the election, the latter data set 2b contains a higher amount of conversational content with a smaller rate of live event comments. A third data set is constituted by the well-known Enron email corpus, containing roughly 250,000 emails exchanged between the employees of the Enron Corporation over 4 years, between Oct 30th 1998 and July 19th 2002. The Enron data is not tied to any specific event, the collapse of the company does not become clearly visible neither in the timelines of volume nor of email length. Data sets 4a and 4b include over 200,000 tweets and 40,000 posts on the online social networking service Facebook (www.facebook.com), respectively, with the common topic of the snow storm "Nemo" which struck the North-eastern coasts of US and Canada on Feb 8th and Feb 9th, 2013. The last data set 5 contains the entire corpus of almost 3 million posts from the Twitter-like microblogging service app.net (www.app.net) over half a year. Similarly to the Enron data set, also here the posts are not tied to any particular events.

The topical focus of the collected messages in several of the data sets is on corresponding ongoing events, which enables us to study the response of individuals and the social collective during these phases of high excitation. Messages before or after events, or messages unrelated to any events provide us with a baseline of no (or low) excitation against which we can compare our measures during phases of high excitation. In the following main text we focus the presentation of our results on the Twitter data set 1, similar or deviating results from the other data sets are reported later and see Figs. S2, S3, S4, S5, S6, and S7 in File S1, stressing the independence of the main result from the medium. Table 1 shows an overview of the key properties of all data sets. The data used in this study are publicly available or can be shared with other researchers upon reasonable request.

### Sampling bias

Particular care has to be taken when sampling from online media such as Twitter. First of all, population in online media is biased towards demographic groups which have easier access to the internet, or may be biased by different interests, in this case interests in Golf or politics, for example. By collecting different data sets from sources with various demographics we alleviate this issue. Second, Twitter users may not always be tied to single human individuals, but e.g. to corporate entities or to automated bots [23]. We examine the issue of selective engagement of user demographics in section "Contraction is not due to selective engagement of specific user demographics".

Further, in the case of Twitter, due to the large amount of data and specific approaches, various methods of extracting data from Twitter can be found in the literature. Some studies base their data on a user-centric approach, in which first a set of users is selected, optionally their alters in the network (and alters of alters through snowball sampling), and then their tweets [13]. Others first select a hashtag or a set of hashtags to fix the topic of interest [24]. In our case we use data in the latter form, i.e. we have a set of hashtags available which puts our focus on the topical content of the collective events. See Ref. [25] for an analysis of the possible ways of selection bias introduced by these methods.

## Results

### Strong anti-correlation between length and rate of messages

We measure and relate the rate of messages, a property of the whole message stream, to message length, a property of the individual messages. The hourly rate of tweets is shown in Fig. 1B. It follows a characteristic diurnal progression, with more tweets during the tournament (days one to five) and substantially less tweets afterwards (day six and later); the dotted line separates these two time frames. The winning move of golfer Bubba Watson on the event's final day triggers a large spike of messages, almost 100,000 tweets during one hour. A large part of these tweets contain the name "Bubba" followed by a varying number of exclamation marks. The content length, measured in number of characters between 1 and 140, is a property associated with each individual tweet, hourly averages are reported in Fig. 1C. We correlate the hourly number $M$ of tweets with the average content length $L$ and find a strong anti-correlation within the time frame of the tournament (Pearson correlation coefficient $\rho = -0.62$, p-value $< 10^{-12}$ for the hypothesis of no correlation). This relation can be reasonably well fit by both a logarithmic form $L \sim \log M$ and a power-law $L \sim M^{\alpha}$ due to the flat slope. The power-law fit is shown in Fig. 1A via green line, the slope is $\alpha \approx -0.077$. Slope and correlation are robust in respect to different time windows, see Figs. S13, S14, and Table S3 in File S1. Volume and content length are not significantly correlated in the low-volume phase after the tournament ($\rho = 0.10$, p-value $= 0.18$). Correlation coefficients obtained in a log-log scale are similar ($\rho = -0.90$, p-value $< 10^{-44}$ during event, $\rho = 0.12$, p-value $= 0.014$ afterwards).

### Lognormal content length distributions

Dividing the probability distribution of content length into logarithmically binned classes of different volumes reveals a smooth transition between the low-volume and high-volume phases showing a wide variety of states, Fig. 2A. In times of low message rates (purple and blue circles), the distribution grows slowly until $L \approx 70$, stays approximately constant until $L \approx 120$ and

**Table 1.** Properties of the data sets ordered by the Pearson correlation coefficient $\rho$ between length $L$ and rate of messages $M$.

| | Data set | $M_{total}$ | Range | Step | Event | $\rho$ | p-value | $f_{\pm 12}$ | $\alpha$ | $\beta$ | $L_{max}$ |
|---|---|---|---|---|---|---|---|---|---|---|---|
| 2a | Forum election | 19,019 | 60h | 10 m | s, o | $-0.73$ | $8.3\times10^{-26}$ | 0.14 | $-0.32$ | $-0.32$ | N/A |
| 1 | Twitter golf | 411,239 | 14d | 1h | s, o | $-0.62$ | $1.6\times10^{-13}$ | 0.42 | $-0.08$ | $-0.12$ | 140 |
| 4b | Facebook snow | 43,516 | 4d | 1h | o | $-0.48$ | $1.2\times10^{-6}$ | – | $-0.21$ | $-0.05$ | N/A |
| 4a | Twitter snow | 219,866 | 4d | 1h | o | $-0.44$ | $8.5\times10^{-6}$ | – | $-0.03$ | $-0.03$ | 140 |
| 2b | Forum pre-election | 10,696 | 8.5d | 10m | o? | $-0.10$ | 0.0008 | – | $-0.04$ | $-0.12$ | N/A |
| 3 | Enron email | 154,003 | 4y | 1d | n | 0.01 | 0.81 | – | 0.08 | 0.01 | N/A |
| 5 | App.net | 2.5 Mio. | 0.5y | 1d | n | 0.10 | 0.20 | – | 0.01 | 0.02 | 256 |

The event type denotes whether an event is unfolding over a non-singular period of time like the (o)ngoing snow storm, if there is an incisive, temporally (s)ingular happening like the winning move in the golf event, or if there is (n)o particular event. Suffixes for ranges and time steps stand for (m)inute, (h)our, (d)ay, (y)ear. The fraction of messages posted in the peak hour, compared with hourly fractions 12 hours prior and afterwards, is denoted by $f_{\pm 12}$ and can be interpreted as the immediacy of the response to singular events. Exponents $\alpha$ and $\beta$ measure the best least-squares fit slopes between $\log L$ and $\log M$, and between $\mu$ and $\log M$, respectively. The symbol $L_{max}$ denotes the imposed length limitation in the respective medium. Although the fraction of peak hour messages is highest for the golf event, correlation is stronger in the presidential election forum, possibly due to the length limitation in data set 1. All other correlations are consistent with the type of event, i.e. correlations are less strong when there is only an ongoing event. We checked for robustness of the parameters in Section S4 in File S1.

peaks shortly before the maximum length of $L=140$. During high-volume phases however (red and orange circles) the distribution grows fast, peaks around $L\approx25$, then decreases to a low local minimum around $L\approx130$ and displays a smaller peak again at $L=140$. Distributions of message lengths in other data sets follow similar shapes. The second peak however is an artifact introduced by message length limitation, therefore it only appears in Twitter and in data set 5.

The functional form of a lognormal $P(\ln L)\sim\mathcal{N}(\mu,\sigma^2)$ serves as a good fit to the distributions of content lengths $L$ from a number of reasonable distributions, see Fig. S8 and Table S1 in File S1. For the Twitter data set fits were performed in the range 1 to 120 since the length limitation of Twitter introduces artificially inflated probabilities close to the maximum length of 140 – users often need to cut back their messages after exceeding the limitation [26], or have their messages automatically truncated or stored externally by apps (for example, by www.twitlonger.com). This effect is absent for media allowing unlimited messages, see Fig. S9, S10, and Table S2 in File S1. We report fits via dashed lines in Fig. 2A, corresponding mean and standard deviation parameters in Fig. 2B and C, detailed fit parameters are given in Table S1 and Table S2 in File S1. The dependence between message rate and the lognormal parameter $\mu$ is clearly decreasing, see dashed line in Fig. 2B. We denote the slope of this curve as $\beta$, values for all data sets are reported in Table 1. The smallest bin, $100\leq M<200$, pink square, contains only values from the time after the event. The standard deviation $\sigma^2$ appears independent of $M$ in the high message rate regime.

## Contraction is not due to selective engagement of specific user demographics

Is it the case that messages contract due to behavioral changes in a fixed population, or does the contraction result from no change in behavior but rather from the engagement of additional, possibly less media-competent, user demographics that affect the behavior of the crowd through their distinct behavioral pattern?

According to the threshold model of collective behavior [27] each individual is modeled to have a threshold of number of peers who must be observed of making a decision before the given individual herself makes the decision too. For example, before you take part in a riot, you might need to see 300 others rioting. The distribution of these thresholds over all individuals may be heterogeneous. In fact, in our case, already the activity rate is highly heterogeneous – the distribution of total tweets per user roughly follows a power law with slope $-2.6$, Fig. 3A. A majority of $66\%$ of users post exactly one tweet, only less than $2\%$ of users tweet more than ten times, being also the main reason why we do not focus on studying the behavior of single users but rather the collective activity. The maximum number of tweets made by a single user is 1619. We define the participation threshold $M_0$ of a user as the smallest hourly message rate in which the user posts a message, Fig. 3B. This definition is based on the threshold model, assuming that the number of tweets is a rough indicator for the feature that drives individuals to become active. As one could expect, some users only tweet during important times of the event – here, for example, a sizable portion of $20\%$ of all users tweet only during the one hour of the tournament's finale. Otherwise, the almost uniform increase of probabilities from high to low values of $M_0$ reflects the heterogeneity of posting behavior over all the different users. The posting volume of single users does not change substantially over time: Fig. 3D shows that the hourly number of tweets $M$ per unique user $U$ is constant around $M/U\approx1.2$ during the event, and slightly higher and noisier afterwards. The threshold $M_0$ averaged over the hourly participating unique users, Fig. 3C, however roughly follows the timeline of volume $M$ meaning that higher volume comes with a larger number of people who only post in such high-volume phases.

Data set 1 in particular contains a mix of message types from a highly heterogeneous set of users, where different effects may be superimposed. For example, if a tweet includes mentions of multiple other usernames, it might be reasonable to assume a conversational nature of the message, e.g. a reply within an ongoing discussion. Moreover, not all accounts may be controlled by single human users, but instead by automated bots or corporate entities which may display different posting behavior. By systematically excluding subsets of messages with such specific properties or which come from particular user groups the observed effects can become clearer or less clear in certain cases. We study the probability distributions of content length $L$ for specific activity classes of users, filtering for number of tweets, for participation thresholds, and several other properties. Figure 4 shows that the effect of message length contraction is clearly visible for all activity classes except for around $6\%$ of tweets of all users who post more than 100 tweets. The strongest peak emerges for those 31.5% of tweets of all users who make exactly one tweet. Further, examining the distributions for various threshold classes of users shows that also users within almost all kinds of threshold ranges show the contraction behavior. Only users with a high threshold, $10^4 < M_0 < 10^5$, are missing most of the contraction since they tend to make single short tweets only, as well as the 4% of tweets of all users with a low threshold $M_0 < 10^2$ which are those who have posts after the event. The number of followers or followees, and the number of hashtags within a tweet, has a similar influence as the number of tweets a user makes. Again, all but the most extreme users show the contraction effect. A strong decline of the contraction effect and its peak is observed in the jump from tweets without mentions to tweets with 1 mention. In conclusion, we see a mix of reasons for the observed phenomena: the contraction phenomenon seems to naturally span over all but the most extreme types of users independent of activity behavior or other properties, but the one-time/high-threshold posters additionally join during high-volume phases and provide an extra boost of short messages. Conversational tweets tend to show less length contraction, as they are presumably less direct reactions to the event.

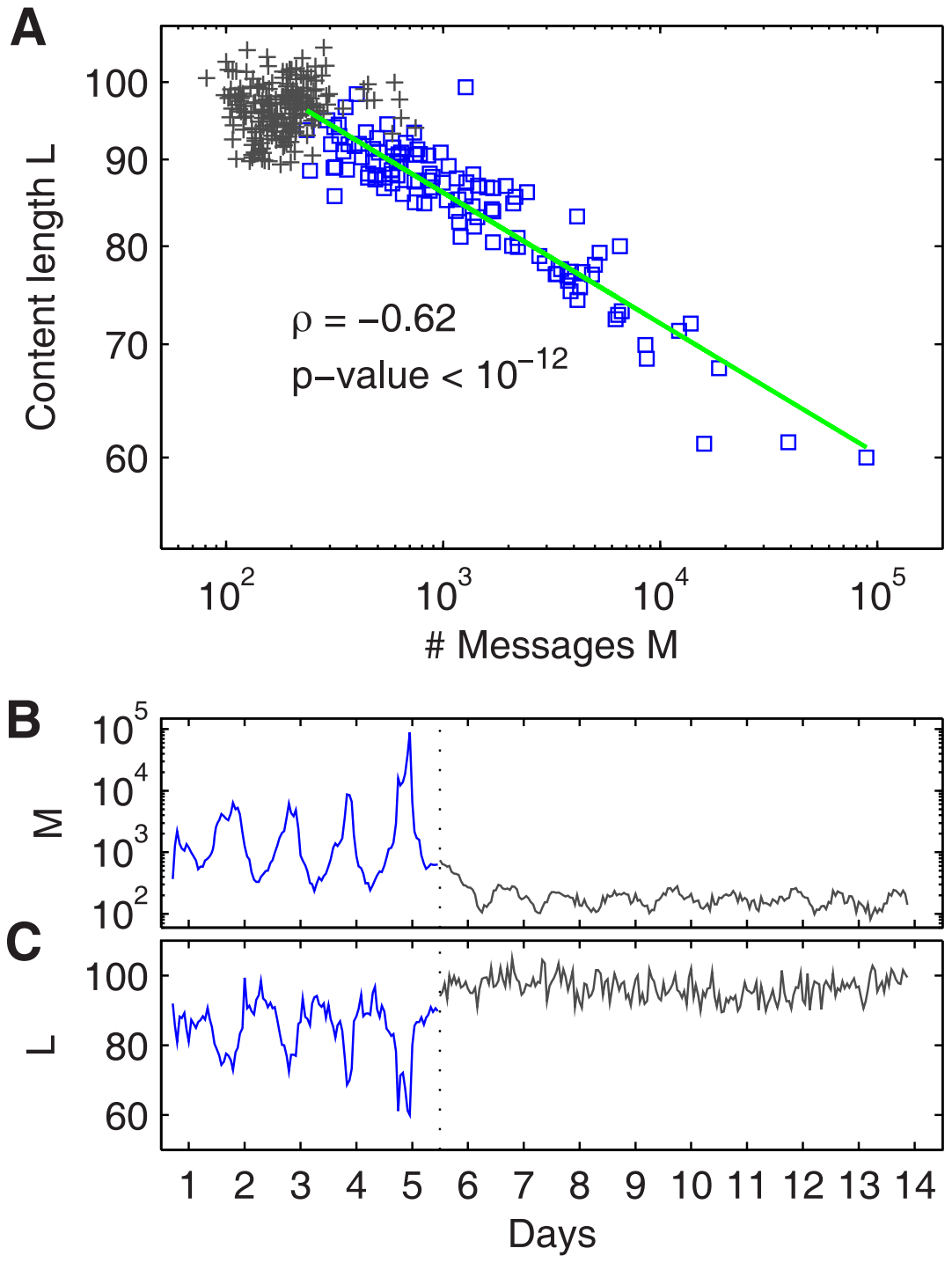

**Figure 1. Strong anti-correlation between length and rate of messages posted during events.** (A) The property of content length $L$ (number of characters per message) can be related to the property of volume $M$ (number of messages per time interval) via power law with slope $\alpha \approx -0.077$ (green line), a logarithmic fit cannot be rejected either due to the flat slope. Blue squares are hourly average values from the first five days in which the Masters Tournament took place, grey crosses are the hourly average values from subsequent times. Message rate and content length are strongly anti-correlated during the Masters ($\rho = -0.62$, p-value $< 10^{-12}$ for the hypothesis of no correlation) but not after the tournament ($\rho = 0.10$, p-value $= 0.18$). (B) Respective message rate and (C) content length over time, averaged hourly. Here all plots refer to the Twitter data set 1, results are similar in other media, see Section S2 in File S1.

### Dependence of correlation strength on event type

In Table 1 we report the different correlation parameters measured for all studied data sets, ordered by decreasing correlation strength $\rho$. Note that here the correlation is measured on the raw data; when the correlation is measured on log-log scales the $\rho$ values are ordered identically, but closer to $-1$ and with much higher significance levels. The "Event" column of Table 1 denotes the type of events occurring. By "o" we denote ongoing events that are unfolding over a period of time, including the 4-day golf tournament (data set 1), the all-night long coverage of the presidential election night (data set 2a), and the snow storm (data sets 4a and 4b) which is predicted earlier and lasts at least for 18 hours, from Friday, Feb 8th afternoon to Saturday, Feb 9th noon. The week-long presidential pre-election thread (data set 2b) might also be classified as an ongoing event. On the other hand we use the letter "s" to depict incisive, temporally singular happenings, as the winning of the golf tournament in the sudden death finale (data set 1) or the confirmation of the winning of the presidential election (data set 2a). Data sets 3 and 5, the Enron email corpus and the app.net posts, do not encompass any specific event we are aware of. In the case of the Enron corpus, although the collapse of the company falls into the considered time range we found no evidence for a particularly abnormal, sudden change in email volume or their contents, apart from a generally decreasing rate of emails in the end of the company's lifetime.

Depending on the type of event, we measure different strengths in the anti-correlation and in the two related exponents $\alpha$ and $\beta$. Singular events are more anti-correlated than non-singular events, the two data sets 3 and 5 which contain no event show no correlation. The exponents also generally follow this trend, however in case of the two Twitter data sets the exponents are much lower. For a possible explanation see Section Discussion below.

Blue bars in Fig. 5A and B show the fractions of messages posted in each hour in the timeframe between 12 hours before and 12 hours after the peak, for the two singular events of the golf tournament finale and the presidential election, respectively. The symbol $f_{\pm 12}$ in Table 1 denotes the fraction of messages posted during the peak hour, for the former event $f_{\pm 12} = 0.42$, for the latter one $f_{\pm 12} = 0.14$. The progression of posts is substantially different from the diurnal patterns during average non-event days, displayed with red bars (we used the daily patterns averaged over the available days after the golf event, for the presidential election thread we used the averaged daily patterns from the pre-election thread, data set 2b). For comparison, the fraction of messages for the non-singular snow storm event has the low value of $f_{\pm 12} = 0.09$ – this is just twice above the value $f_{\pm 12} = 1/24 \approx 0.042$ which would be expected from a uniform posting behavior. Thus, here the posting pattern is not substantially deviating from the natural daily pattern on the timescale of 24 hours. However, Fig. 5C shows that the event still becomes highly visible if the timescale is shifted from hours to days.

## Discussion

### Anti-correlation as a proxy for collective arousal

Generally, we measure a stronger anti-correlation between length and rate of messages the more temporally singular the event is, see $\rho$ in Table 1. Clearly, $\rho$ and p-values are lowest for the two singular events, closer to zero but still negative for the ongoing events, and not significant for the data sets in which no particular event is unfolding. Together with the assumption that emotional messages are shorter, this observation suggests that the strength of anti-correlation might be a good proxy for the "collective arousal" of an event, i.e. how big of an emotional response the event causes. (Using sentiment analysis for validation purposes seems not feasible due to the typically very short messages during high volume phases.) We stress that our measurements cannot be interpreted as explicit evidence for the contraction of messages due to emotions – this is rather a reasonable assumption we make here that has to be checked in future research. An alternative suggestion was made in a previous study which also observed that the length of tweets tends to decrease during important events [24], but without referring to emotional influences. There, it was suggested that users could have a possible increased attentiveness to the event and a desire to spend less time typing.

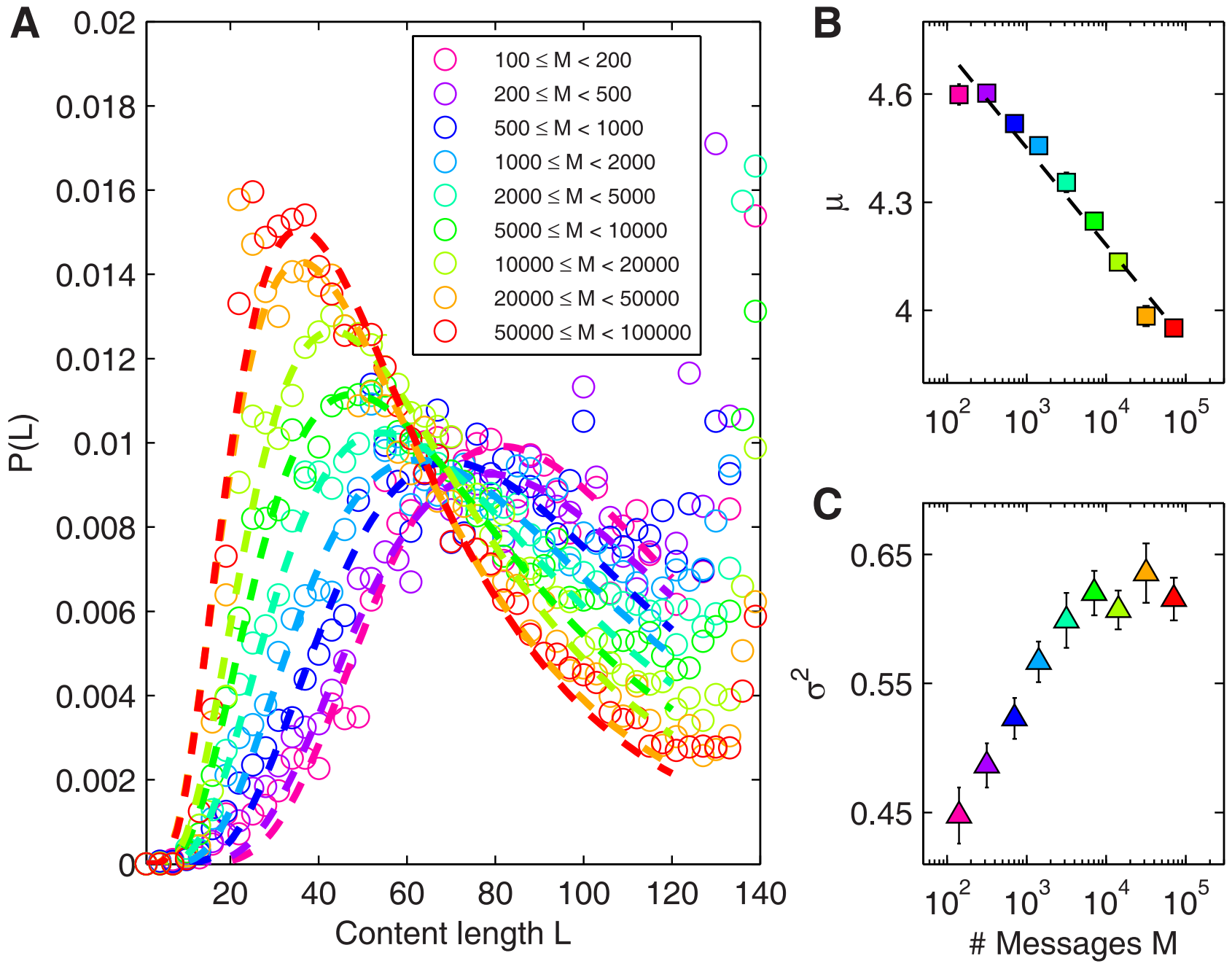

**Figure 2. Lognormal distribution of message lengths and dependence of its parameters on excitation.** (A) Probability distributions of content length $L$ of messages gathered by logarithmically binned classes of different hourly volume $M$ (circles), and corresponding lognormal fits (dashed curves, fit ranges 0 to 120). During low-volume phases (pink and blue), the distribution grows slowly. For high-volume phases (orange and red) however the distribution grows fast and peaks at $L \approx 25$. Peaks at the maximum length of 140 are an artifact from the length limitation in the specific medium (Twitter), absent for unlimited media, see Section S2 in File S1. For visual clarity only every third data point is shown. (B) Plot of the lognormal fit parameter $\mu$ against message rate $M$ demonstrates the systematic relation between message rate and length, dashed line. (C) Plot of the lognormal fit parameter $\sigma^2$ versus message rate $M$. Here the value of $\sigma^2$ increases with the message rate $M$ to some point and appears independent of the volume class in high volume regimes. Error bars denote 95% confidence intervals.

Nevertheless, some issues of parameter sensitivity have to be discussed with care. First, the correlation coefficient $\rho$ shows sensitivity to the time step used, while the exponents $\alpha$ and $\beta$ are more robust, see Table S3 in File S1. Second, despite the golf event having the by far largest value of $f_{\pm 12}$ and therefore features the most temporally singular response, its exponent $\alpha = -0.08$ is only the third smallest after the Forum election with $\alpha = -0.32$ and the Facebook snow storm response with $\alpha = -0.21$. A straightforward explanation for this difference is the limitation $L_{\max} = 140$ in Twitter, which severely limits the possible slopes: Note that the y-axis in Fig. 2 spans less than half a decade, while data sets from unlimited media span one or more decades, see Section S2 in File S1. Similarly, we find that the exponent $\beta$ is smallest for the Forum election data with $\beta = -0.32$, and only second for the Twitter event with $\beta = -0.12$. Other slopes are in line with the ordering, except for the slope $\beta = -0.12$ for the pre-election forum thread. In this case the small number of four data points for determining $\beta$ may introduce inaccuracies however, see Fig. S3 in File S1.

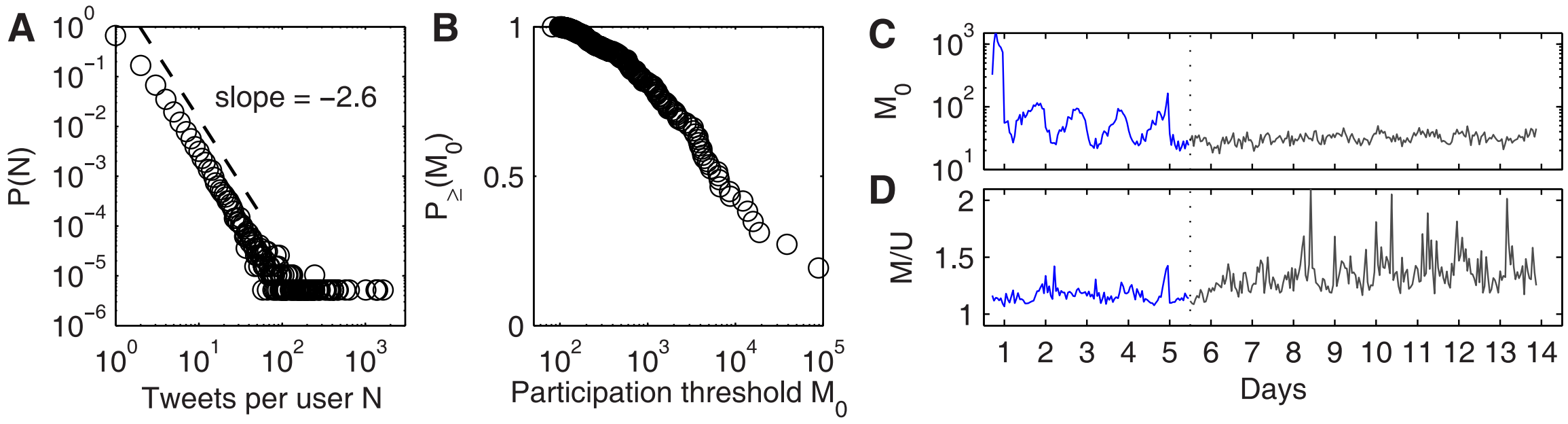

**Figure 3. Distributions of individual activity features and their timelines.** (A) Distribution of number of tweets per user $N$ during the whole time span. The distribution follows approximately a power law with slope $-2.6$. (B) Cumulative distribution of absolute thresholds $M_0$ of all users, i.e. the smallest hourly volume for each user in which a tweet is posted. Approximately 20% of users only post during the one hour which marks the final of the event, but roughly one fourth of the users also post during hours in which less than 1000 tweets are posted. (C) Participation thresholds $M_0$ over time ($M_0$ is measured separately for each user over the whole timespan; for each hour we average the $M_0$ values of all unique users who tweet in that hour). The curve follows roughly the volume curve of $M$, showing that high volume phases feature additional users who only post during those phases. (D) Timeline of number of tweets $M$ per unique user $U$, $M/U$. During the event, each user posts on average around 1.2 tweets per hour, with a particular peak at the finale of the tournament, showing that single users write slightly more messages during that time of high excitation. After the event, the individual activity increases slightly due to the departure of the masses of casual Twitter users.

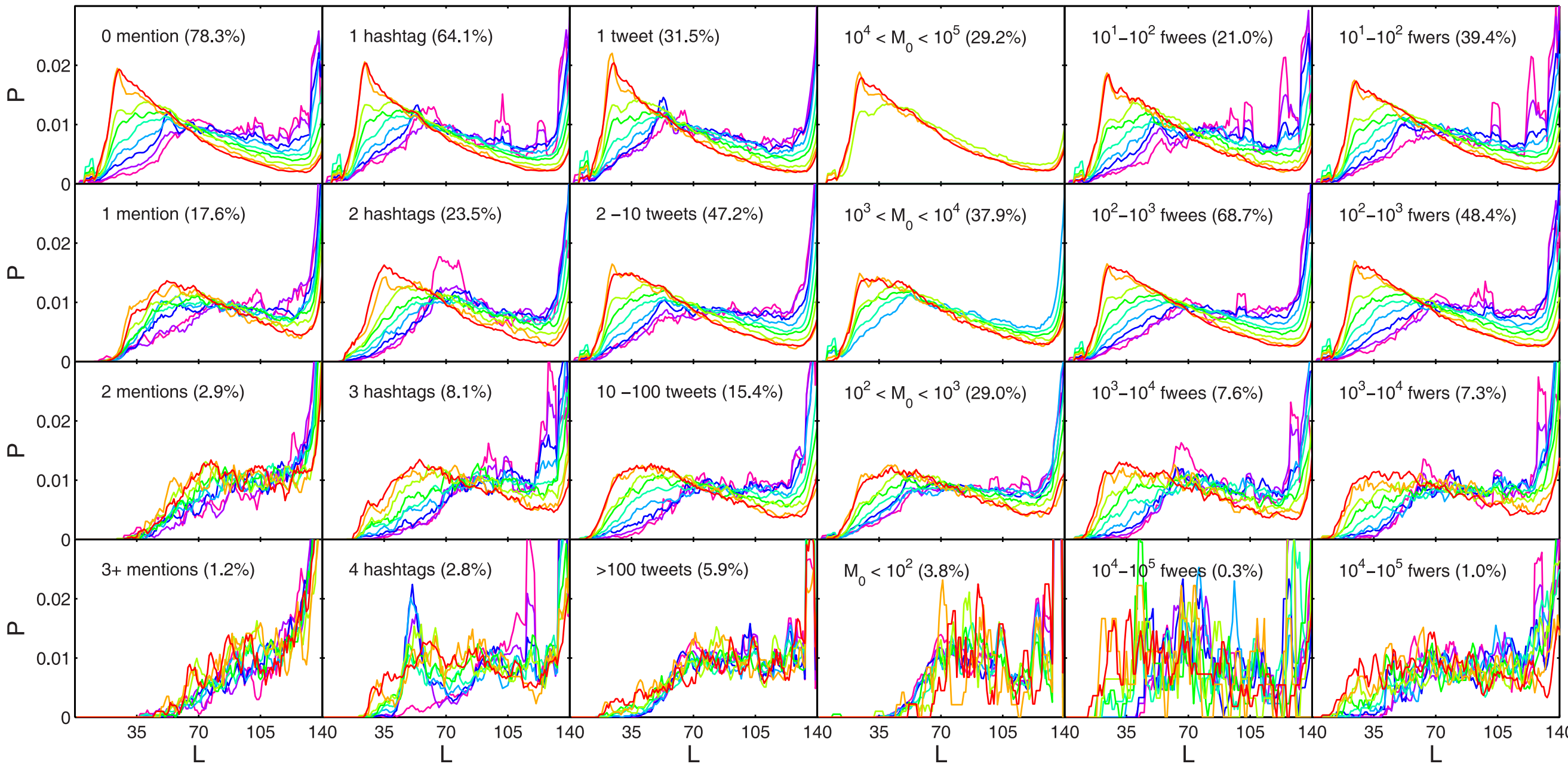

**Figure 4. Message length distributions filtered for classes of tweets with certain properties.** Colors correspond to different volume bins, see legend in Fig 2A. The percentage shown in each subplot corresponds to the percentage of tweets within the data set matching the corresponding criterion. The contraction effect appears throughout all classes, except for users in the highest percentiles of activity. Tweets of users who post less than others or who have less followers or followees than others, or which contain no mentions indicating that the message is not of conversational nature, show the strongest contraction. Users with a high threshold, $10^4 < M_0 < 10^5$, are missing most of the contraction since they tend to make single short tweets only. For visual clarity we applied a moving average filter of length 5 to all curves.

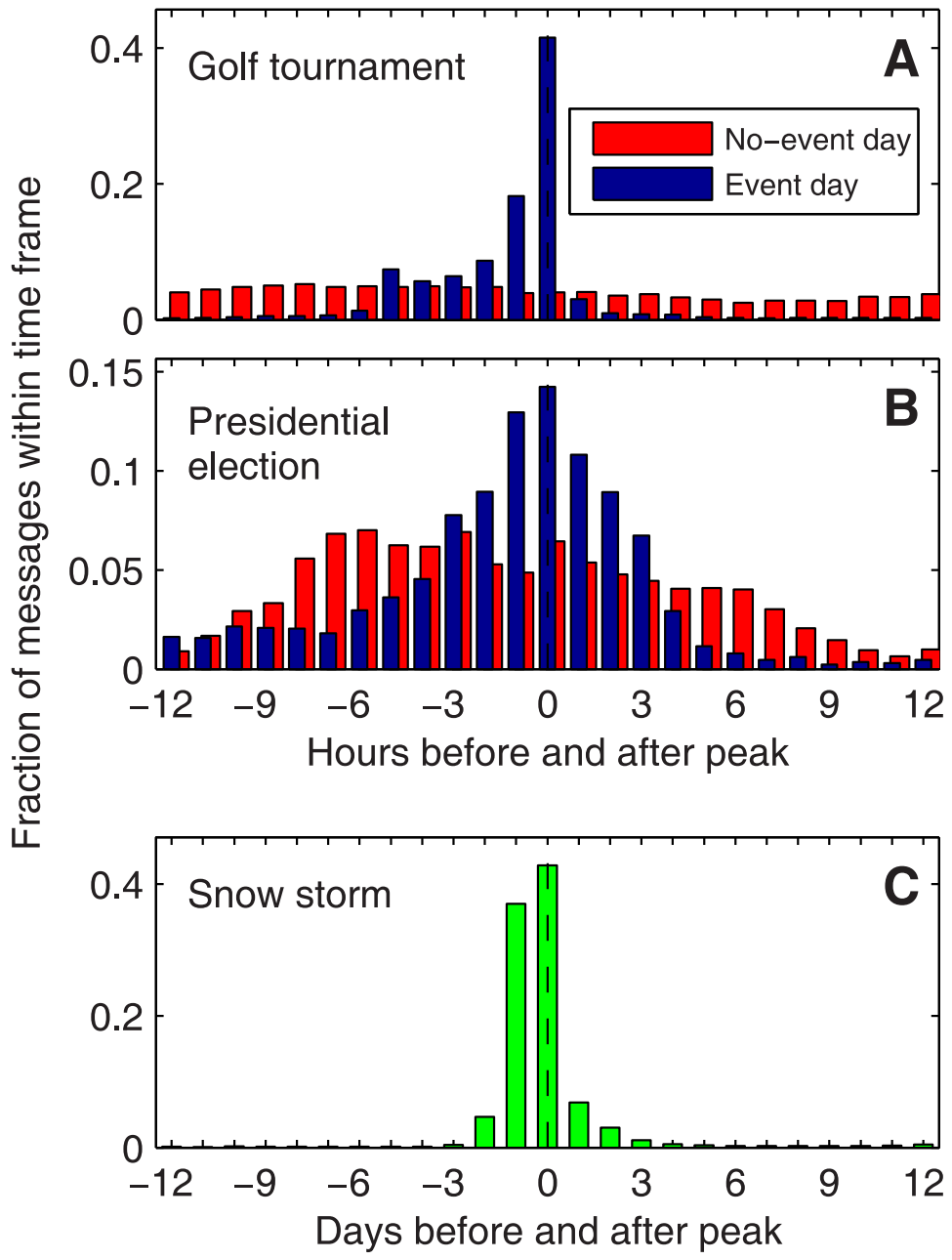

**Figure 5. Fraction of messages posted immediately before and after an event peak.** The fraction during the peak hour serves as a measure for the temporal singularity of the event. (A) The largest peak occurs during the finale of the golf tournament, data set 1, with a peak fraction of $f_{\pm 12h} = 0.42$. (B) The second singular event is the presidential election thread, data set 2a, with a peak fraction of $f_{\pm 12h} = 0.14$. (C) The snow storm event, data set 4a, on the other hand, shows a clear peak only when the time scale is coarsened, from hours to days. The snow storm is thus "much less singular" than the previous events, but still a distinct event in the broader view.

Although they have a time of anticipation, both the golf event and the presidential election are exogenous, since the outcome of the event (who is the winner?) is unknown beforehand, and because they do not arise because of, say, endogenous viral propagation of a meme in the social network [17]. We have no data set of response to exogenous events available where there is no time of anticipation, such as a tsunami or an earthquake, but due to their exogenous nature we presume that the effect of systematic message length contraction could be observed as well in that case.

## Phenomenological response laws

Although message length limitations that are in place in particular media may distort some results and have to be accounted for on a case by case basis, in general our analysis suggests that all media are equivalent in the way their users respond to events. To stress this point, we propose here a unifying framework to describe the collective behavior of different online media users, independent of the medium used. Summing up our previous insights, we found that the length $L$ of a message can be related to the ongoing message rate $M$ through two phenomenological laws. First,

$$\langle L(M) \rangle \sim M^{\alpha}, \tag{1}$$

where $\alpha < 0$ measures the strength of the anti-correlation between $\langle L \rangle$ and $M$. Second, the lognormal distribution

$$P(\ln L) \sim \mathcal{N}(\mu(M), \sigma^2(M)), \tag{2}$$

whose mean and standard deviation parameters depend on the message rate $M$ as follows:

$$\mu(M) \equiv \langle \ln L(M) \rangle = \mu_0 + \beta \ln M = \ln(M/M_0)^{\beta} \tag{3}$$

$$\sigma^2(M) \simeq \sigma^2_{\infty} \tag{4}$$

Parameters $\mu_0$ and $M_0 = \exp(-\mu_0)$ are two constants corresponding to a reference state, and $\sigma_{\infty}$ is a typical parameter characterizing the dispersion of individual responses in 'ideal cases' (as defined in Section S4 in File S1). The constant relation given by Eq. (4) is verified in ideal cases, see Figs. S11 and S12 in File S1, and in high message rate regimes, but is less clear in other cases. Eq. (3), which is always clearly verified, interestingly provides a link between the collective – through the aggregated measure $M$ – and the individual – through the random variable $L$ – responses to the ongoing event. This link is embodied by the exponent $\beta < 0$ which is a media specific measure of the singularity of an event.

At this point, it is worth noting that the lognormal distribution modeled in Eq. (2) is not unexpected. Lognormal distributions are supported by works in Stylometry, the statistical study of linguistic style, on general sentence length distributions [28–31]. They have also been found in a variety of other areas such as file sizes or durations of telephone calls and seem to be a universal phenomenon [32]. In any case, the distribution of sentence lengths in particular being relatively understudied leaves both origin and validity of the lognormal distribution an open question.

Since they measure the same phenomenon, one can expect the two phenomenological laws to be related. One property of the lognormal distribution is that $\langle L(M) \rangle = \exp(\mu + \sigma^2/2)$, which, combining Eqs (3) and (4), leads to $\langle L(M) \rangle = (M/M_0)^{\beta} \exp(\sigma^2_{\infty}/2) \sim M^{\beta}$. Compatibility between Eqs (1) and (2) thus imposes the equality $\alpha = \beta$, which is well backed by our measurements in case of purely singular events (datasets 1 and 2a, see Table 1). This relation is not respected in case of non-singular events, where $\sigma^2$ fluctuates and Eq. (4) does not hold.

## A response-to-stimuli relationship

As stated before, Psychophysics aims at understanding how the subjective sensation $\psi$ of an individual is related to the stimuli $S$ he is subjected to, and in turn how the individual response $R$ is related to his sensation $\psi$. Empirical laws describing such relationships in the context of physical stimuli (such as sound, light, temperature, etc.) have traditionally been proposed based on the analysis of the responses of individuals. In the same spirit, the two phenomenological laws given by Eqs. (1) and (2) can be seen as first steps towards a psychophysic science of online collective behavior.

Let us stress that due to the nature of the available online data, we lack precise observation on a number of levels. Several processes are uncertain here: Is the response based on a single-origin stimulus, e.g. are all the users reacting to the same mass medium broadcast, or are there collective or network effects where users are driven by messages of other users and not always directly by the event itself? How much do these possible network effects depend on the particular online medium? Previous work on youtube views suggests that both effects can play a dominant role [17]. In that case, a power-law distributed memory kernel $\phi(t)$ was used to describe the waiting time of an individual to view a video after a time $t$ that she was subjected to the cause (the cause may include any reason such as chance or a newspaper entry). In our case we are however dealing with messages instead of views, and with online media in which typically events are either commented on live or discussed prior or subsequently to their time of occurrence (except of course for the cases in which no particular events are commented on). The explicit response to a singular event we therefore assume as instantaneous, while messages before or after the event we assume to stem from anticipation, discussions, or from other ongoing events, all of which may be hard to untangle. Further, we can reasonably assume that the driving force for the increase in arousal is the event itself or the associated "hype" from news media or in the social networks of the users, and we disregard secondary effects such as the circadian arousal rhythm [33].

Being fully aware of the limitation of our observations, we nevertheless propose a loose analogy between our phenomenological laws and psychophysics models. Indeed, the message rate $M$ being an indicator of the number of people following an event, Fig. 3D, which in turn is linked to the intrinsic interest of the event and to the probability that you hear about that event through some of your online contacts, it can be taken as a proxy for the stimulus $S$ the users are subjected to. The collective response $R$ can be measured through $\langle L \rangle$. Lacking any measure on the emotional sensation of the individual, let us assume a simple relationship between the response $R$ and the sensation $\psi$ in the form $R = \langle L \rangle = \psi^{-1}$. Under these assumptions, we could write a relation $\psi = \langle L \rangle^{-1} = M^{-\alpha} = S^{\gamma}$, with $\gamma = -\alpha > 0$, meaning that the suggested phenomenological law of Eq. (1) can be understood as a collective analogy to Stevens law [22].

## Assessment of possible dissatisfaction from imposed message length limitation

A curious side-effect of the almost perfectly followed lognormal distribution we find in unlimited media is the possibility to assess possible user dissatisfaction from truncation effects in media which allow only limited messages. From its very inception, Twitter faced complaints about its "trademark", the arbitrary character length limitation of $L_{\max} = 140$ created originally for compatibility with SMS messaging [34,35]. A truncation of messages leads to a deviation of the empirical density function from the theoretical lognormal probability density function $f_{\mathrm{logn}}(x;\mu,\sigma) = 1/\left(x\sigma\sqrt{2\pi}\right) e^{-\frac{(\ln x - \mu)^2}{2\sigma^2}}$, where the probability mass above $L_{\max}$ is packed into the range below the threshold. Therefore, the mass of the fitted lognormal pdf which lies above the threshold, $p_{\mathrm{diss}} = \int_{L_{\max}}^{\infty} f_{\mathrm{logn}}(x;\mu,\sigma)\mathrm{d}x$, gives a good approximation of the percentage of messages that had to be truncated, i.e. the possibly dissatisfying to compose messages. During the special high volume phases ($50{,}000 < M < 10{,}000$), $p_{\mathrm{diss}}$ is at 5%, however, during the low volume phases ($100 < M < 200$), $p_{\mathrm{diss}}$ grows to a substantial proportion of 22%. Raising the truncation threshold from 140 to $L_{\max} = 170$ by only 30 characters already halves $p_{\mathrm{diss}}$ to 11%, another raise to $L_{\max} = 200$ yields a further halving to $p_{\mathrm{diss}} \approx 6\%$. In comparison, applying app.net's threshold of $L_{\max} = 256$ results in the relatively negligible proportion of $p_{\mathrm{diss}} \approx 1.7\%$, which is 13 times lower than the $p_{\mathrm{diss}}$ determined for Twitter.

## Conclusion

In this work we contributed to the understanding of the regularities of online response to collective events using different online media. We investigated online messages from various media created in response to major, collectively followed events, and found a medium-independent, systematic relation between content length and message rate, which is amplified by aroused followers who are getting excited by the event. Further, we identify the distributions of content lengths as lognormals in line with statistical linguistics, and suggest a phenomenological law for the systematic dependence of the message rate on the lognormal mean parameter. Our method could be re-applied to other datasets, extended to offline forms of communication like call durations of mobile phone users, where evidence of lognormals already exists [36], or to economic transactions. The exact mechanism which leads to the lognormal distribution in the context of text length is still an ongoing debate. Current research is based on collection of texts or sentences without discriminating between the excitement of the authors. Our findings may provide new ingredients for the development of future models incorporating this aspect, and might have possible applications in marketing or spam detection.

Finally, the observation of substantial truncation effects may have implications for user experience policies in the design of mobile or online micro-blogging and messaging services. In the case of the existing service Twitter, an increase of the message limit by only 30 characters halves the number of situations where users want to post a tweet longer than 140 characters but are constrained by the system. Some of these situations might correspond to cases that cause dissatisfaction to users, suggesting that the current tradeoff between data storage and user experience could be improvable by an increase of message length limits. However, such a "relief" for some of the composing users could come with an impairment of the ability of users who follow many people to process the incoming content and to engage with it. In this context, a number of psychological questions of user experience need to be addressed in future research with empirical data on the user level before coming to conclusive assessments.

## Supporting Information

**File S1 Supporting Information. Figure S1,** Characteristic of data set 1 with a one hour timestep. **Figure S2,** Characteristics of data set 2a: Forum thread during the presidential election night. **Figure S3,** Characteristics of data set 2b: Forum thread one week before the presidential election night. **Figure S4,** Characteristics of data set 3: Enron email corpus. **Figure S5,** Characteristics of data set 4a: Twitter snow storm. **Figure S6,** Characteristics of data set 4b: Facebook snow storm. **Figure S7,** Characteristics of data set 5: App.net. **Figure S8,** Length distribution of all the tweets in data set 1, and corresponding fits of various distributions. **Figure S9,** Length distribution of all the posts in data set 2a, and corresponding fits of various distributions. **Figure S10,** Detail from Fig. S9 showing the tail of the distribution. **Figure S11,** Dependence of the lognormal fitted parameters with the volume. **Figure S12,** Analysis of the`ideal' subset of tweets. **Figure S13,** Characteristic of data set 1 with a 5 min timestep. **Figure S14,** Characteristic of data set 1 with a 3 hours timestep. **Table S1,** Best fit MSEs ($\times 10^6$) for the tested distributions for the Twitter data set 1. **Table S2,** Best fit MSEs ($\times 10^6$) for the tested distributions for the forum data set 2a. **Table S3,** Characteristic parameters measured on data set 1 for different timesteps.
(PDF)

## Acknowledgments

We thank Stewart Townsend (Datasift), Moritz Stefaner and Stephan Thiel (Studio NAND) for providing data set 1 and corresponding background information, and Markus Schläpfer, Roberta Sinatra, and Stanislav Sobolevsky for stimulating discussions. Sebastian Grauwin acknowledges generous support from Ericsson. We would further like thank the National Science Foundation, the AT&T Foundation, the Rockefeller Foundation, the MIT SMART program, the MIT CCES program, Audi Volkswagen, BBVA, The Coca Cola Company, Ericsson, Expo 2015, Ferrovial, GE and all the members of the MIT Senseable City Lab Consortium for supporting this research.

## Author Contributions

Conceived and designed the experiments: MS SG CR. Performed the experiments: MS SG. Analyzed the data: MS SG. Contributed reagents/materials/analysis tools: MS SG. Wrote the paper: MS SG.

## References

1. Lazer D, Pentland A, Adamic L, Aral S, Barabási AL, et al. (2009) Computational social science. Science 323: 721.
2. Kleinberg J (2003) Bursty and hierarchical structure in streams. Data Mining and Knowledge Discovery 7: 373–397.
3. Barabási AL (2005) The origin of bursts and heavy tails in human dynamics. Nature 435: 207–211.
4. Klimek P, Bayer W, Thurner S (2011) The blogosphere as an excitable social medium: Richter's and Omori's law in media coverage. Physica A 390: 3870–3875.
5. Leskovec J, Backstrom L, Kleinberg J (2009) Meme-tracking and the dynamics of the news cycle. In: Proceedings of the 15th ACM SIGKDD international conference on Knowledge discovery and data mining. ACM, pp. 497–506.
6. Szell M, Thurner S (2010) Measuring social dynamics in a massive multiplayer online game. Social Networks 32: 313–329.
7. Szell M, Lambiotte R, Thurner S (2010) Multirelational organization of large-scale social networks in an online world. Proceedings of the National Academy of Sciences 107: 13636–13641.
8. Szell M, Thurner S (2012) Social dynamics in a large-scale online game. Advances in Complex Systems 15: 1250064.
9. Asur S, Huberman B, Szabo G, Wang C (2011) Trends in social media: Persistence and decay. In: Proceedings of the Fifth International AAAI Conference on Weblogs and Social Media (ICWSM'11).
10. Yang J, Leskovec J (2011) Patterns of temporal variation in online media. In: Proceedings of the fourth ACM international conference on Web search and data mining. ACM, pp. 177–186.
11. Lehmann J, Gonçalves B, Ramasco J, Cattuto C (2012) Dynamical classes of collective attention in twitter. In: Proceedings of the 21st international conference on World Wide Web. ACM, pp. 251–260.
12. Bagrow J, Wang D, Barabási A (2011) Collective response of human populations to large-scale emergencies. PloS one 6: e17680.
13. Grabowicz P, Ramasco J, Moro E, Pujol J, Eguiluz V (2012) Social features of online networks: The strength of intermediary ties in online social media. PLoS ONE 7: e29358.
14. Golder SA, Macy MW (2011) Diurnal and seasonal mood vary with work, sleep, and daylength across diverse cultures. Science 333: 1878–1881.

15. Thurner S, Szell M, Sinatra R (2012) Emergence of good conduct, scaling and zipf laws in human behavioral sequences in an online world. PLoS ONE 7: e29796.
16. Klimek P, Thurner S (2013) Triadic closure dynamics drives scaling laws in social multiplex networks. New Journal of Physics 15: 063008.
17. Crane R, Sornette D (2008) Robust dynamic classes revealed by measuring the response function of a social system. Proceedings of the National Academy of Sciences 105: 15649.
18. Head H (1920) Aphasia and kindred disorders of speech. Brain 43: 87–165.
19. Paus T (2000) Functional anatomy of arousal and attention systems in the human brain. Progress in brain research 126: 65–77.
20. Weber D (1846) Der Tastsinn und das Gemeingefühl. Handwörterbuch der Physiologie III 549: 1846.
21. Fechner G (1860) Elemente der Psychophysik. Leipzig, Breitkopf & Härtel 1.
22. Stevens S (1957) On the psychophysical law. Psychological review 64: 153.
23. Tavares G, Faisal A (2013) Scaling-laws of human broadcast communication enable distinction between human, corporate and robot twitter users. PLOS ONE 8: e65774.
24. Shamma D, Kennedy L, Churchill E (2010) Tweetgeist: Can the twitter timeline reveal the structure of broadcast events. CSCW Horizons.
25. González-Bailón S, Wang N, Rivero A, Borge-Holthoefer J, Moreno Y (2012) Assessing the bias in communication networks sampled from twitter. Available at SSRN 2185134.
26. Boyd D, Golder S, Lotan G (2010) Tweet, tweet, retweet: Conversational aspects of retweeting on twitter. In: System Sciences (HICSS), 2010 43rd Hawaii International Conference on. IEEE, pp. 1–10.
27. Granovetter M (1978) Threshold models of collective behavior. American journal of sociology : 1420–1443.
28. Williams C (1940) A note on the statistical analysis of sentence-length as a criterion of literary style. Biometrika 31: 356–361.
29. Wake W (1957) Sentence-length distributions of Greek authors. Journal of the Royal Statistical Society Series A (General) 120: 331–346.
30. Holmes D (1985) The analysis of literary style–a review. Journal of the Royal Statistical Society Series A (General) : 328–341.
31. Furuhashi S, Hayakawa Y (2012) Lognormality of the distribution of japanese sentence lengths. Journal of the Physical Society of Japan 81: 034004.
32. Sobkowicz P, Thelwall M, Buckley K, Paltaglou AG, Sobkowicz A (2013) Lognormal distributions of user post lengths in internet discussions - a consequence of the weber-fechner law? EPJ Data Science 2.
33. Dickman SJ (2002) Dimensions of arousal: Wakefulness and vigor. Human Factors: The Journal of the Human Factors and Ergonomics Society 44: 429–442.
34. Sarno D (2009) Twitter creator Jack Dorsey illuminates the site's founding document. Part I. The *Los Angeles Times*. Available: http://latimesblogs.latimes.com/technology/2009/02/twitter-creator.html. Accessed 2013 Aug 15.
35. Milian M (2009) Why text messages are limited to 160 characters. The *Los Angeles Times*. Available: http://latimesblogs.latimes.com/technology/2009/05/invented-text-messaging.html. Accessed 2013 Aug 15.
36. Guo J, Liu F, Zhu Z (2007) Estimate the call duration distribution parameters in gsm system based on kl divergence method. In: Wireless Communications, Networking and Mobile Computing, 2007. WiCom 2007. International Conference on. IEEE, pp. 2988–2991.